\documentclass[twocolumn]{aastex631}

\usepackage{graphicx}
\usepackage{txfonts}
\usepackage{multirow}
\usepackage{hyperref}
\usepackage{mathrsfs}
\usepackage{xcolor}

\shorttitle{Modeling the Chromosphere and Transition Region of Planet-hosting Star GJ~436}
\shortauthors{Hintz et al.}

\begin{document}

\title{Modeling the Chromosphere and Transition Region of Planet-hosting Star GJ~436
       }

\author[0000-0002-5274-2589]{Dominik~Hintz}
\affiliation{University of Arizona, Lunar and Planetary Laboratory,
             1629 E University Boulevard, Tucson, AZ 85721, USA;
             \href{mailto:dhintz@lpl.arizona.edu}{dhintz@lpl.arizona.edu}
             }
\affiliation{Hamburger Sternwarte, University of Hamburg,
             Gojenbergsweg 112, D-21029 Hamburg, Germany}

\author[0000-0002-1046-025X]{Sarah~Peacock}
\affiliation{University of Maryland, Baltimore County, MD 21250, USA}
\affiliation{NASA Goddard Space Flight Center, Greenbelt, MD 20771, USA
             }

\author[0000-0002-7129-3002]{Travis~Barman}
\affiliation{University of Arizona, Lunar and Planetary Laboratory,
             1629 E University Boulevard, Tucson, AZ 85721, USA;
             \href{mailto:dhintz@lpl.arizona.edu}{dhintz@lpl.arizona.edu}
             }

\author{Birgit~Fuhrmeister}
\affiliation{Hamburger Sternwarte, University of Hamburg,
             Gojenbergsweg 112, D-21029 Hamburg, Germany}

\author[0000-0002-4019-3631]{Evangelos~Nagel}
\affiliation{Institut für Astrophysik und Geophysik,
             Friedrich-Hund-Platz 1, D-37077 G\"ottingen, Germany}

\author[0000-0002-1624-0389]{Andreas~Schweitzer}
\affiliation{Hamburger Sternwarte, University of Hamburg,
             Gojenbergsweg 112, D-21029 Hamburg, Germany}

\author{Sandra~V.~Jeffers}
\affiliation{Max-Planck-Institut für Sonnensystemforschung,
             Justus-von-Liebig-Weg 3, D-37077 G\"ottingen, Germany}

\author[0000-0002-6689-0312]{Ignasi~Ribas}
\affiliation{Institut de Ci\`encies de l'Espai (ICE, CSIC),
             - Campus UAB, c/ de Can Magrans s/n, E-08193 Bellaterra, Barcelona, Spain}
\affiliation{Institut d'Estudis Espacials de Catalunya (IEEC), E-08034 Barcelona, Spain}

\author[0000-0003-1242-5922]{Ansgar~Reiners}
\affiliation{Institut für Astrophysik und Geophysik,
             Friedrich-Hund-Platz 1, D-37077 G\"ottingen, Germany}

\author{Andreas~Quirrenbach}
\affiliation{Landessternwarte, Zentrum f\"ur Astronomie der Universit\"at Heidelberg,
             K\"onigstuhl 12, D-69117 Heidelberg, Germany}

\author[0000-0002-8388-6040]{Pedro~J.~Amado}
\affiliation{Instituto de Astrof\'isica de Andaluc\'ia (CSIC),
             Glorieta de la Astronom\'ia s/n, E-18008 Granada, Spain}

\author[0000-0002-5086-4232]{Victor J. S. B\'{e}jar}
\affiliation{Instituto de Astrofisica de Canarias, c/ Via Lactea s/n, E-38205 La Laguna, Tenerife, Spain}
\affiliation{Departamento de Astrofsica, Universidad de La Laguna, E-38206 Tenerife, Spain }

\author[0000-0002-7349-1387]{Jos\'{e} A. Caballero}
\affiliation{Centro de Astrobiolog\'{i}a, CSIC-INTA, Camino Bajo del Castillo s/n,
             E-28692 Villanueva de la Ca\~nada, Madrid, Spain}

\author{Artie P. Hatzes}
\affiliation{Th\"uringer Landessternwarte Tautenburg, Sternwarte 5, D-07778 Tautenburg, Germany }

\author[0000-0002-7779-238X]{David Montes}
\affiliation{Departamento de F{\'i}sica de la Tierra y Astrof{\'i}sica \&
IPARCOS-UCM (Instituto de F\'{i}sica de Part\'{i}culas y del Cosmos de la UCM),
\newline
Facultad de Ciencias F{\'i}sicas, Universidad Complutense de Madrid, E-28040 Madrid, Spain}

\accepted{June 20, 2023}
\submitjournal{ApJ}

\begin{abstract}
  Ahead of upcoming space missions intending
  to conduct observations of
  low-mass stars
  in the ultraviolet (UV) spectral region
  it becomes imperative to simultaneously conduct atmospheric modeling
  from the UV to the visible (VIS) and near-infrared (NIR).
  Investigations on extended spectral regions will help to improve the
  overall understanding of the diversity of spectral lines
  arising from very different atmospheric temperature regions.
  Here we investigate atmosphere models with a
  chromosphere and transition region for the M2.5V star GJ~436,
  which hosts a close-in Hot Neptune.
  The atmosphere models are
  guided by observed spectral features from the UV to the VIS/NIR
  originating in the chromosphere and transition region of GJ~436.
  High-resolution observations from the Hubble Space Telescope
  and Calar Alto high-Resolution search for M dwarfs with Exo-earths with
  Near-infrared and optical Echelle Spectrographs (CARMENES)
  are used to obtain an appropriate model spectrum for the investigated M~dwarf.
  We use a large set of atomic species considered in nonlocal
  thermodynamic equilibrium conditions within our PHOENIX model computations
  to approximate the physics within the low-density atmospheric regions.
  In order to obtain an overall match for the
  nonsimultaneous observations,
  it is necessary to apply a linear combination of two model spectra,
  where one of them better reproduces the
  UV lines
  while the other better represents the lines
  from the VIS/NIR range.
  This is needed to adequately handle different activity states
  across the observations.
\end{abstract}

\keywords{stars: activity -- stars: chromospheres -- stars: transition region
          -- stars: individual (GJ 436) -- stars: low mass }

\section{Introduction}
  Modeling the lowest-density regions of stellar atmospheres,
  from the chromosphere to corona
  (also referred to as the upper atmosphere in the following),
  remains a challenging exercise.
  Processes such as
  acoustic heating \citep[][]{Wedemeyer2004A&A...414.1121W},
  ambipolar diffusion, and conduction
  \citep[][]{Fontenla1990ApJ...355..700F,Fontenla1991ApJ...377..712F,Fontenla1993ApJ...406..319F}
  are believed to contribute to the nonradiative heating
  in the upper atmosphere, but incorporating these into a
  self-consistent atmosphere model remains difficult given
  the range of densities and temperatures.
  Using semiempirical one-dimensional temperature structures,
  as first used for modeling different regions on the surface
  of the Sun \citep[][]{Vernazza1973,Vernazza1981}, turned out
  to be highly effective in computing synthetic spectra
  that match spectral lines arising from its upper atmosphere.
  Atmosphere codes such as PHOENIX
  \citep{Hauschildt1992JQSRT..47..433H, Hauschildt1993JQSRT..50..301H, Hauschildt1999JCoAM.109...41H}
  can take into account the upper atmosphere, i.~e., the chromosphere and transition region
  (designating the steep rise in temperature between chromosphere and corona), and
  have been proven capable of reproducing observed
  chromosphere and transition region lines of M-dwarf stars from the ultraviolet (UV)
  \citep[e.~g.,][]{Peacock2019ApJ...871..235P, Peacock2019ApJ...886...77P}
  to the visible (VIS) and near-infrared (NIR) wavelength ranges
  \citep[e.~g.,][]{Hintz2019A&A...623A.136H, Hintz2020A&A...638A.115H}.

  The M2.5V star GJ~436 and its closely orbiting
  Hot Neptune GJ~436b have been continuously in the
  spotlight of scientific investigations throughout the last two
  decades.
  GJ~436 is located 9.78\,pc away from the Sun
  \citep{Gaia2021refId0} and has a radius of 0.42\,$R_\odot$
  and a mass of 0.44\,$M_{\odot}$ \citep{Rosenthal2021ApJS..255....8R}.
  This M-dwarf star exhibits periodic variations on short- and long-term scales.
  The rotational period of
  44.6~days \citep{DA2019A&A...621A.126D}
  and an activity cycle of about 7.4~yr
  \citep{Lothringer2018AJ....155...66L}
  have been detected from photometric variations.
  Chromospheric activity indicators have been found to vary on similar timescales
  due to rotation \citep{Mascareno2015MNRAS.452.2745S} and the activity cycle
  \citep{Kumar2023MNRAS.518.3147K}.
  GJ~436b orbits its host star at 0.03\,au with an orbital period of 2.64\,days
  \citep[][]{Butler2004ApJ...617..580B} and it is well known that
  high-energy radiation, from the host star
  in the UV and X-ray range, can significantly
  influence the atmosphere of
  exoplanets \citep[e.~g.,][]{Lammer2007AsBio...7..185L, Tian2008JGRE..113.5008T}.
  Due to the close orbit of GJ~436b, the high-energy radiation from GJ~436
  strongly influences the Hot Neptune's atmosphere
  and drives its atmospheric evolution
  \citep{Kulow2014ApJ...786..132K,Ehrenreich2015Natur.522..459E,Bourrier2016A&A...591A.121B}.
  From extensive and continued observations of GJ~436
  and access to precise transmission spectroscopy of GJ~436b,
  it becomes essential to study the low-density
  part of the atmosphere of the host star
  in order to better understand the characteristics of GJ~436
  and, eventually, to improve our understanding of GJ~436b.

  Stellar activity is a common property of M-dwarf stars
  and there are a number of spectral lines that are sensitive to this activity
  as well as to the temperatures across the upper atmospheric layers
  \citep[][]{Vernazza1981,Fontenla1990ApJ...355..700F,Andretta1997ApJ...489..375A,Fuhrmeister2005A&A...439.1137F}.
  In particular, the hydrogen H$\alpha$ line
  at 6564.6\,$\AA$ (vacuum wavelength), and
  located in the visible part of the spectrum,
  is a valuable and often-used activity indicator
  \citep[e.~g.,][]{Stauffer1986ApJS...61..531S,Gizis2002AJ....123.3356G,Robertson2016ApJ...832..112R}.
  However, this line exhibits
  an ambiguous behavior in the low-activity regime, i.~e.,
  with increasing stellar activity,
  the absorption depth increases before filling in and reversing
  into an emission line with
  even stronger activity states \citep[][]{Cram_Mullan_1979ApJ...234..579C}.
  This complicated behavior can lead to misinterpretation, and so
  various chromospheric lines should be considered
  when studying stellar activity and related atmospheric properties.
  The chromospheric lines of
  the \ion{Ca}{2} infrared triplet (IRT)
  ($\lambda\lambda =$ 8500.4, 8544.4, 8664.5\,$\AA$) and
  \ion{Na}{1}~D doublet ($\lambda\lambda =$ 5891.6, 5897.6\,$\AA$) constitute other
  tracers used for activity-related investigations
  \citep[e.~g.,][]{GomesDaSilva2011,Martinez-Arnaiz2011MNRAS.414.2629M,Robertson2016ApJ...832..112R,Martin2017A&A...605A.113M}.
  While the H$\alpha$ line is formed in the upper chromosphere,
  the \ion{Ca}{2}~IRT and \ion{Na}{1}~D lines
  can be used to characterize the lower chromosphere
  \citep[e.~g.,][]{Hintz2019A&A...623A.136H}.
  The \ion{He}{1} IRT lines
  ($\lambda\lambda =$ 10\,832.1, 10\,833.2, 10\,833.3\,$\AA$)
  are sensitive to temperatures
  from around 10\,000 to 20\,000~K and therefore are
  good probes of the base of the transition region
  \citep[e.~g.,][]{Andretta1995ApJ...439..405A,Andretta1997ApJ...489..375A,Hintz2020A&A...638A.115H}.
  Their line formation is actually dependent on the nonlocal UV continuum
  shortward of 504~$\AA$ which is capable of ionizing the neutral helium in the first place
  before the line's lower energy level can be populated by electrons
  \citep[e.~g.,][]{Andretta1995ApJ...439..405A, Andretta1997ApJ...489..375A, Hintz2020A&A...638A.115H}.
  This highlights the necessity of further extending
  stellar atmosphere modeling by taking into account broad wavelength ranges
  from the UV to the NIR region.
  The \ion{He}{1} IRT lines are also often used
  to measure and investigate the atmospheres of exoplanets
  while transiting their host stars
  \citep[][]{Allart2018Sci...362.1384A,Nortmann2018Sci...362.1388N,Salz2018A&A...620A..97S,Spake2018Natur.557...68S}.

  The hydrogen Ly$\alpha$ line ($\lambda =$ 1215.7\,$\AA$),
  which contributes a significant portion of flux to
  the UV part of the spectrum, is used to study the
  shape of the transition region and considered a proxy
  for the often inaccessible extreme UV (EUV) wavelength range
  \citep[e.~g.,][]{Linsky2014ApJ...780...61L,Peacock2019ApJ...871..235P}.
  Furthermore, irradiation from Ly$\alpha$
  is also known to influence the composition of planetary atmospheres
  \citep[e.~g.,][]{Vidal-Madjar2004ApJ...604L..69V,Trainer2006PNAS..10318035T}.
  The \ion{C}{4} lines at 1550\,$\AA$ are also relatively
  strong lines originating from the transition region
  and are formed at even higher temperatures than
  Ly$\alpha$ \citep[][]{Linsky2017ARA&A..55..159L}.
  Both of these lines have previously been used to
  model the transition region
  \citep[e.~g.,][]{Peacock2019ApJ...871..235P,Peacock2019ApJ...886...77P}.
  The lines discussed above are all sensitive to different atmospheric
  regions and thus will provide the empirical guidance we need to construct
  a model atmosphere capable of reproducing the overall UV through IR spectrum of GJ436.

  GJ~436 has already been the subject of stellar atmosphere modeling
  in the UV \citep{Peacock2019ApJ...886...77P} as well as in the
  VIS and NIR \citep{Hintz2019A&A...623A.136H,Hintz2020A&A...638A.115H}
  wavelength regions.
  Both studies used the PHOENIX code to obtain
  appropriate models to reproduce the spectral features considered.
  These studies modeled the
  UV and VIS/NIR lines seperately, the work of this paper
  models these regions simultaneously
  to obtain an overall match to the observations.

  Section \ref{SEC_Phx_model} gives an overview about
  how the PHOENIX models are constructed
  and important model properties.
  Thereupon, we compare our synthetic spectra
  to GJ~436 observations (Section~\ref{SEC_comp_to_obs})
  obtained from the Hubble Space Telescope
  (HST) for the UV region and from the
  Calar Alto high-Resolution search for M dwarfs with Exo-earths with
  Near-infrared and optical Echelle Spectrographs
  \citep[CARMENES;][]{Quirrenbach2018SPIE10702E..0WQ}
  for the VIS/NIR range.
  Finally, in Section~\ref{SEC_sum_concl} we summarize our
  results and give a conclusion of this work.

\section{Model construction} \label{SEC_Phx_model}
  For our modeling of GJ~436,
  we use the stellar atmosphere code PHOENIX
  \citep{Hauschildt1992JQSRT..47..433H, Hauschildt1993JQSRT..50..301H, Hauschildt1999JCoAM.109...41H}.
  The PHOENIX code has the capability to
  compute the synthetic spectrum based
  on a given semiempirical, ad hoc 1D temperature structure
  \citep{Fuhrmeister2005A&A...439.1137F} where we are
  able to specify the temperature structure
  of the chromosphere and transition region
  and include a nonlocal thermodynamic equilibrium (NLTE)
  treatment of many atomic and ion species.
  The PHOENIX code is still the subject of further ongoing improvements:
  for instance the work by \citet[][]{Peacock2019ApJ...871..235P, Peacock2019ApJ...886...77P}
  recently added a partial frequency redistribution mode
  \citep[][]{Hubeny1995ApJ...455..376H,Uitenbroek2001ApJ...557..389U}
  in order to better match observed resonance lines such as the Ly$\alpha$ line
  in the UV spectra of a few M~dwarfs.
  This makes the PHOENIX code a good choice
  to create model spectra for spectral lines that are observed in late-type stars
  and form in the upper atmospheres of these stars.

  By introducing free parameters that determine the atmospheric position
  of the temperature minimum or the top of the chromosphere
  \citep[see, e.~g.,][for a detailed description]{Hintz2019A&A...623A.136H,Peacock2019ApJ...871..235P},
  we construct such a semiempirical, ad hoc 1D temperature structure
  representing the chromosphere and transition region.
  On top of the basic photosphere model (with the stellar parameters
  of GJ~436 as given in Table~\ref{tab_gj436_parameters})
  we attach an atmospheric structure with temperatures increasing outwards
  up to $\sim$200\,000~K.

\begin{figure}[t]
\includegraphics[width=0.47\textwidth]{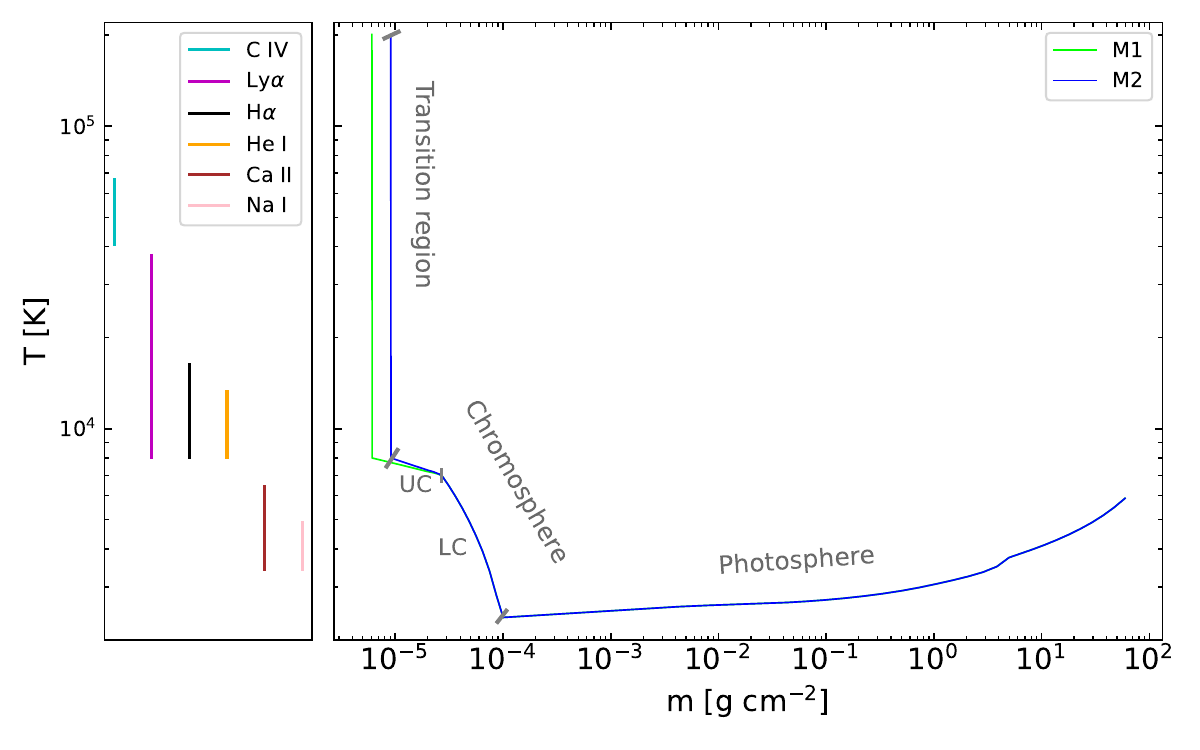}
\caption{
    Temperature profiles of models M1 (green) and M2 (blue)
    as a function of column mass density (right panel).
    On the left-hand side, we indicate approximate atmospheric temperatures
    above the temperature minimum
    where most of the line fluxes of Ly$\alpha$,
    \ion{C}{4}, H$\alpha$, \ion{He}{1} IRT,
    \ion{Ca}{2} IRT, and \ion{Na}{1}~D$_2$
    in model M2 originate from
    according to the flux contribution function
    \citep[][]{Magain1986A&A...163..135M,Fuhrmeister2006A&A...452.1083F,Hintz2019A&A...623A.136H}.
    The corresponding temperature regions
    of M1 are similar. For the \ion{Ca}{2} IRT and \ion{Na}{1}~D$_2$ lines,
    they are actually the same, while they are slightly lower for the other lines,
    but $\sim$2500\,K at most.
    For M2 we indicate by gray lines the boundaries
    of the transition region and upper (UC) and lower chromosphere (LC),
    as well as the photosphere.
    }
\label{fig_T_struct}
\end{figure}

  In Fig~\ref{fig_T_struct}, we show two example
  atmosphere temperature structures we considered in this paper.
  The depicted models M1 and M2 differ by the onset of the transition
  region in terms of the column mass density -- M1's transition region is located
  at $m=-5.2\,$dex while for M2 it is $m=-5.0\,$dex.
  In Fig~\ref{fig_T_struct}, we also indicate the temperature regions of M2
  where the majority of flux emerges for spectral lines
  discussed in this work.
  These temperature ranges were determined from the flux contribution function
  \citep[][]{Magain1986A&A...163..135M,Fuhrmeister2006A&A...452.1083F}
  and following the method described by \citet{Hintz2019A&A...623A.136H}.
  A line formation analysis is beyond the scope of this paper
  \citep[for this, see, e.~g.,][]{Falchi1998A&A...336..281F};
  however, with this we can get an impression
  about the temperature
  regions responsible in the chromosphere and transition region.
  The onset of the transition region of model M2 is located at
  a density almost $\sim$1.6 times larger
  than in the case of model M1.

  Using different increases in temperature at different atmospheric heights
  follows a well-known approach already
  tested and applied to M~dwarfs by several former model
  studies of the Sun and M~dwarfs
  \citep[e.~g.,][]{Vernazza1981,Andretta1997ApJ...489..375A,Fuhrmeister2005A&A...439.1137F}.
  Furthermore, the general shape of this temperature structure is similar to the one
  obtained for the Sun by balancing radiative losses and the nonradiative
  heating mechanisms of thermal conduction and ambipolar diffusion as introduced by
  \citet{Fontenla1990ApJ...355..700F, Fontenla1991ApJ...377..712F, Fontenla1993ApJ...406..319F}.
  However, the steep temperature gradient of the transition region
  simulates the conduction mostly responsible for the energy transport
  at these heights \citep{Fuhrmeister2005A&A...439.1137F}
  and turned out to be an appropriate model approximation up to a temperature of
  200\,000\,K \citep{Peacock2019ApJ...871..235P,Peacock2019ApJ...886...77P}.

\begin{table*}[h!t]
\caption{
Spectral observations of GJ~436 used
for the considered spectral features.
}
\label{tab_gj436_obs}
\begin{center}
\footnotesize
\begin{tabular}{l c c c c r}
\hline\hline
Spectral Feature  &  Wavelengths [$\AA$]  &  Spectrograph  &  Average Resolving Power  &  Observation Date &   References \\
\hline
 Ly${\alpha}$ & 1215.7 & HST/STIS & 14\,400 & 2015 Jun 24 & 1, 2, 3 \\
 \ion{C}{4} & 1548.2 & HST/COS & 16\,500 & 2015 Jun 25 & 1, 2, 3 \\
 NUV & 2000--3000 & HST/STIS & 3000 & 2015 Jun 24 & 1, 2, 3 \\
 \ion{Na}{1} D$_2$ & 5891.6 & CARMENES/VIS & 94\,600 & 2016 Jan 8 - 2016 Nov 20 & 4 \\
 H$\alpha$ & 6564.6 & CARMENES/VIS & 94\,600 & 2016 Jan 8 - 2016 Nov 20  & 4 \\
 \ion{Ca}{2} IRT & 8500.4 & CARMENES/VIS & 94\,600 & 2016 Jan 8 - 2016 Nov 20 & 4 \\
 \ion{He}{1} IRT & 10\,833.3 & CARMENES/NIR & 80\,400 & 2016 Jan 8 - 2016 Nov 20  & 4 \\
\hline
\end{tabular}
\end{center}
\tablecomments{
We use vacuum wavelengths throughout the paper.
STIS: Space Telescope Imaging Spectrograph; COS: Cosmic Origins Spectrograph; NUV: near-UV.}
\tablerefs{
(1) \citet{France2016ApJ...820...89F};
(2) \citet{Youngblood2016ApJ...824..101Y};
(3) \citet{Loyd2016ApJ...824..102L};
(4) \citet{Quirrenbach2018SPIE10702E..0WQ}.
}
\end{table*}

\begin{table}[h!t]
\caption{
Stellar parameters of GJ~436.
}
\label{tab_gj436_parameters}
\begin{center}
\footnotesize
\begin{tabular}{l c r}
\hline\hline
 Parameter                 &    Value             &   Reference \\
\hline
 M$_{\star}$ [M$_{\odot}$] & 0.441 $\pm$ 0.009      &   1 \\
 R$_{\star}$ [R$_{\odot}$] & 0.417 $\pm$ 0.008      &   1 \\
 d [pc]                    & 9.775 $\pm$ 0.003      &   2 \\
 SpT                       & M2.5                  &   3 \\
 T$_\mathrm{eff}$ [K]             & 3533 $\pm$ 26          &   4 \\
 logg [dex]                & 4.83                 &   4 \\
 v$_\mathrm{rad}$ [km/s]          & 9.59 $\pm$ 0.001       &   5 \\
\hline
\end{tabular}
\end{center}
\tablerefs{
(1) \citet{Rosenthal2021ApJS..255....8R};
(2) \citet{Gaia2021refId0};
(3) \citet{Alonso-Floriano2015};
(4) \citet{Marfil2021A&A...656A.162M};
(5) \citet{Fouque2018MNRAS.475.1960F}.
}
\end{table}

  Recent studies of M-dwarf stars using
  such 1D ad hoc temperature structures
  in the upper atmosphere were able to
  adequately match chromospheric and transition region lines of
  H$\alpha$, \ion{Ca}{2}~IRT, \ion{Na}{1}~D, and \ion{He}{1}~IRT
  in the VIS and NIR wavelength ranges in high-resolution spectra
  \citep{Hintz2019A&A...623A.136H, Hintz2020A&A...638A.115H}.
  Furthermore, such models could also be used to
  successfully reproduce spectral lines of Ly$\alpha$ and \ion{Mg}{2}
  in the UV wavelength range of M-dwarf stars
  \citep{Peacock2019ApJ...871..235P, Peacock2019ApJ...886...77P}.

  In the upper atmosphere where the temperature
  strongly increases and the atmosphere becomes optically thin,
  it is necessary to account for NLTE conditions
  when treating the most abundant atomic and ion species.
  Therefore, the following elements are computed in NLTE:
  H~\textsc{i}, He~\textsc{i}-\textsc{ii}, C~\textsc{i}-\textsc{iv},
  N~\textsc{i}-\textsc{v}, O~\textsc{i}-\textsc{v},
  Ne~\textsc{i}-\textsc{ii}, Na~\textsc{i}-\textsc{iii},
  Mg~\textsc{i}-\textsc{iv}, Al~\textsc{i}-\textsc{iv},
  Si~\textsc{i}-\textsc{iv}, P~\textsc{i}-\textsc{ii},
  S~\textsc{i}-\textsc{iii}, Cl~\textsc{i}-\textsc{iii},
  Ar~\textsc{i}-\textsc{iii}, K~\textsc{i}-\textsc{iii},
  Ca~\textsc{i}-\textsc{iii}, Ti~\textsc{i}-\textsc{iv},
  V~\textsc{i}-\textsc{iii}, Cr~\textsc{i}-\textsc{iii},
  Mn~\textsc{i}-\textsc{iii}, Fe~\textsc{i}-\textsc{iv},
  Co~\textsc{i}-\textsc{iii}, and Ni~\textsc{i}-\textsc{iii}.
  We treat a total of 74 species in NLTE; for
  most of them we use level data provided by PHOENIX
  \citep{Kurucz1995all..book.....K}, but for He~\textsc{i}-\textsc{ii}
  and \ion{C}{4} we use the CHIANTI V7 \citep{Landi2013ApJ...763...86L} database.
  As suggested by \citet{Fuhrmeister2006A&A...452.1083F},
  we do not include LTE background lines when treating hydrogen in NLTE.

  In addition to electron collisions
  we employ hydrogen collisional rates according
  to the empirical findings of
  \citet{Drawin1968ZPhy..211..404D,Drawin1969ZPhy..225..470D},
  which are important for the formation of line wings
  where hydrogen densities
  are orders of magnitude higher than electron densities
  \citep{Barklem2007A&A...466..327B}.
  This particularly accounts for a correct
  treatment of the atmospheric region around
  the temperature minimum where collisions are generally still
  important but the electron density is comparably low.
  Further details about the model construction
  are available in \citet{Hintz2019A&A...623A.136H}
  and \citet{Peacock2019ApJ...871..235P}.

\section{Comparison to observations} \label{SEC_comp_to_obs}
  In this work we use spectral observations from the
  HST taken from
  the Measurements of the Ultraviolet Spectral Characteristics of
  Low-mass Exoplanetary Systems Treasury Survey
  \citep[MUSCLES;][]{France2016ApJ...820...89F,Youngblood2016ApJ...824..101Y,Loyd2016ApJ...824..102L}
  to analyze UV lines of GJ~436.
  For investigating the VIS and NIR region of GJ~436,
  we use spectral observations from CARMENES
  \citep[][]{Quirrenbach2018SPIE10702E..0WQ}.
  Due to a high level of telluric contamination in the
  NIR region around the \ion{He}{1}~IRT lines
  we use the CARMENES coadded template spectrum obtained
  from the CARMENES reduction pipeline \citep{Zechmeister2018A&A...609A..12Z,Nagel2020phd}.
  The considered HST observations were taken in 2015 June
  while the CARMENES observations cover the period
  from 2016 January to 2016 November.

  Investigating possible long-term cycles within the CARMENES data,
  \citet{Fuhrmeister2023A&A...670A..71F} did not find any
  activity-related trend for GJ~436. Yet, apart from CARMENES data, they
  could detect an $R'_\mathrm{HK}$ long-term cycle over a period of more than 17~yr
  with variations of less than $\sim$20\,\% within the corresponding
  sinusoidal best fit of the $R'_\mathrm{HK}$ modulation.
  \citet{Kumar2023MNRAS.518.3147K} reported long-term periodicities
  of H$\alpha$ and \ion{Na}{1} indicators of $\sim$6~yr, which is in line
  with the detected photometric cycle
  \citep{Lothringer2018AJ....155...66L,Loyd2023AJ....165..146L}.
  The activity study by \citet{Hintz2019A&A...623A.136H}
  included GJ~436, but could not find significant variations
  within the investigated chromospheric lines from the activity state.
  The time period of the observational data used
  from CARMENES only covers $\sim$1/6 of the reported H$\alpha$ and
  \ion{Na}{1} indicator perodicities and
  $\sim$1/17 of the detected $R'_\mathrm{HK}$ cycle.
  Therefore, we only use a small fraction of any of these and do not
  expect significantly strong variations within the time period
  of 2016 we use for our coadded CARMENES spectrum.
  Furthermore, we checked the considered lines in the coadded spectrum
  from CARMENES in timespans of $\sim3$ and $\sim6$ months for
  variations in 2016 and could not find significant changes, either.
  Thus, we consider it reasonable to use
  the averaged VIS/NIR CARMENES spectrum for our investigation.

  We convolve our PHOENIX spectra to the spectral resolutions
  of the instruments in the respective wavelength ranges.
  HST spectra have a resolving power at the wavelengths of
  the Ly$\alpha$ and \ion{C}{4} lines of $\sim$14\,400
  and $\sim$16\,500 respectively,
  while the resolving power is $\sim$94\,600 in the CARMENES VIS channel
  (covering \ion{Na}{1}~D, H$\alpha$, \ion{Ca}{2}~IRT)
  and $\sim$80\,400 in the NIR channel (\ion{He}{1}~IRT).
  Table~\ref{tab_gj436_obs} provides a summary
  of the considered spectral features
  and respective observations of GJ~436 used in this work.
  An overview of the stellar parameters of GJ~436
  is given in Table~\ref{tab_gj436_parameters}.

\begin{figure}[h!t]
\includegraphics[width=0.47\textwidth]{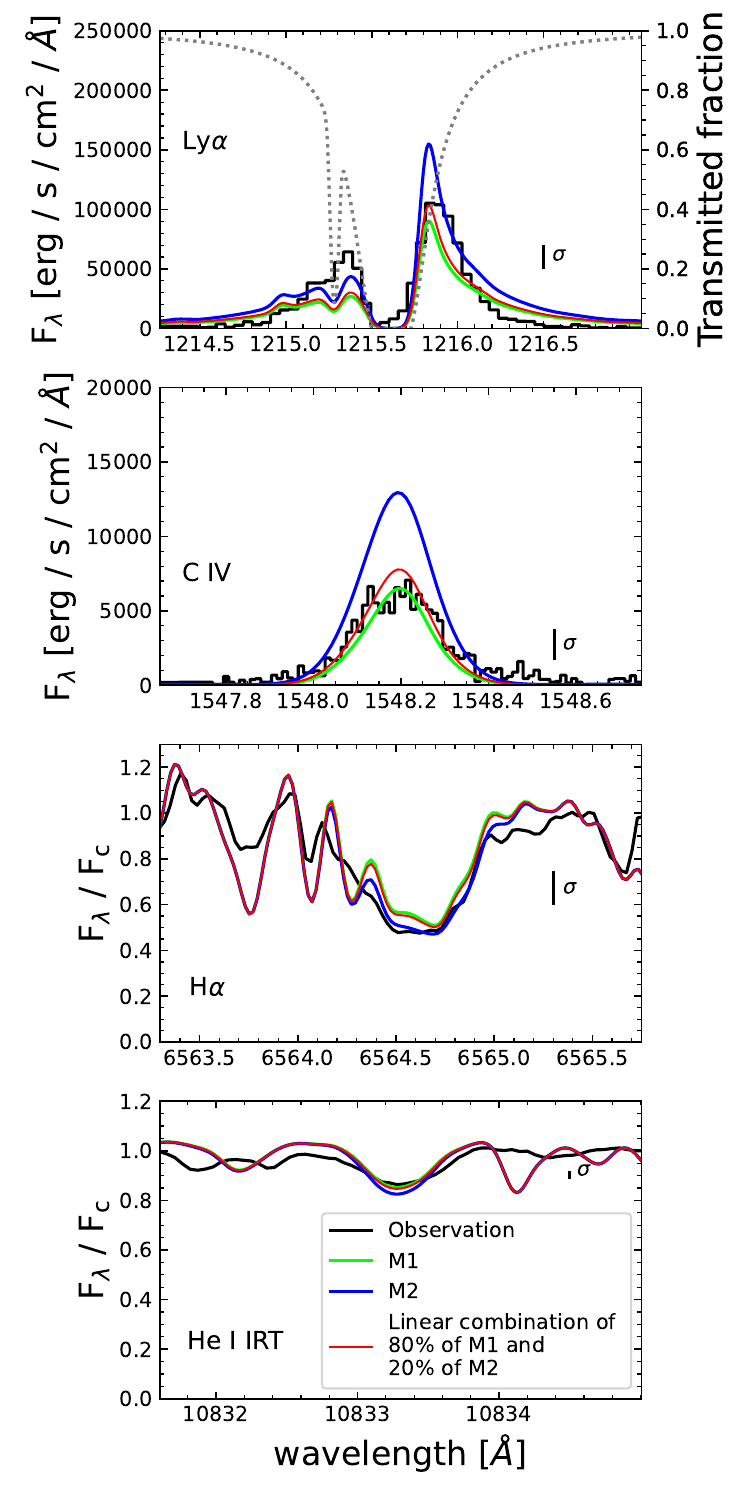}
\caption{
    Models M1 and M2 (green and blue lines) obtained from the temperature structures
    depicted in Fig~\ref{fig_T_struct} (with corresponding color-coding)
    and observations (black lines) of GJ~436 shown in the lines of
    Ly$\alpha$ together with the
    ISM transmittance curve for GJ~436 provided by
    \citet[][gray dashed line; an exemplary insight into how we apply the
    ISM transmission to our model M2 is provided in
    Appendix~\ref{SEC_App_ISM}]{Youngblood2016ApJ...824..101Y}
    (\textit{top panel}),
    \ion{C}{4} (\textit{second panel}),
    H$\alpha$ (\textit{third panel}),
    \ion{He}{1}~IRT (\textit{bottom panel}).
    Red lines correspond to the best linear
    combinations of the model components M1 and M2.
    HST observations are scaled by d$^2$/R$_{\star}^2$
    to the stellar surface. VIS/NIR lines of CARMENES observations
    and model spectra are continuum-normalized.
    The $\sigma$-values correspond to the standard deviations
    of the observed line flux
    (for Ly$\alpha$ $\sigma$ is calculated from the line wings only).
}
\label{fig_lines_comp}
\end{figure}

\subsection{Single models}
  In order to find a chromospheric PHOENIX model in good
  agreement with UV and VIS/NIR observations of GJ~436,
  we use two spectral lines observed by HST
  representing the UV region as well as two lines
  from the VIS/NIR range observed by CARMENES.
  Therefore, we are using two lines
  to represent each wavelength region.
  The UV region is here represented by the lines of
  hydrogen Ly$\alpha$ at $1215.7\,\AA$ and
  \ion{C}{4} at $1548.2\,\AA$.
  When comparing our synthetic spectra to the observed
  Ly$\alpha$ line of GJ~436, we apply the interstellar medium
  (ISM) transmission curve as found by \citet{Youngblood2016ApJ...824..101Y}
  in order to compare our modeled Ly$\alpha$ line to
  what HST is actually able to observe
  with respect to the interstellar absorption along the
  line of sight.
  For the VIS/NIR wavelength range, we use two of the
  activity indicators also applied by \citet[][]{Hintz2020A&A...638A.115H}
  to find a matching model:
  the hydrogen H$\alpha$ line at $6564.6\,\AA$,
  and the strongest component of the \ion{He}{1}~IRT at $10\,833.3\,\AA$
  which consists of the two red overlapping
  triplet lines. The blue component of the triplet at
  $10\,832.1\,\AA$ is not observable for this spectral type,
  but there may be stronger molecular lines
  in this wavelength region \citep{Fuhrmeister2019A&A...632A..24F}.

  In Figure~\ref{fig_lines_comp},
  we show the observed lines as well as the modeled lines obtained from
  the atmosphere structures depicted in Figure~\ref{fig_T_struct}.
  For the CARMENES observations, we normalize both the observations
  and model spectra to the local continua since
  the CARMENES spectra are not flux-calibrated.
  While the synthetic spectrum of model M1 better reproduces the
  \ion{C}{4} line ($\chi^2_\mathrm{C\,IV,\mathrm{M1}} = 9.7$
  compared to $\chi^2_\mathrm{C\,IV,\mathrm{M2}} = 103.7$),
  M2 yields a better match for H$\alpha$
  ($\chi^2_\mathrm{H\alpha,\mathrm{M1}} = 12.8$
  compared to $\chi^2_\mathrm{H\alpha,\mathrm{M2}} = 5.1$).
  The situation in Ly$\alpha$ and the \ion{He}{1}~IRT is less clear, i.~e.,
  $\chi^2$-values of M1 and M2 differ by a factor of $\lesssim2$ for these lines.
  Model M2
  overpredicts \ion{C}{4} and the right Ly$\alpha$ wing,
  but matches H$\alpha$ and also the \ion{He}{1}~IRT.
  On the other hand, M1 fits \ion{C}{4} and the \ion{He}{1}~IRT,
  but underpredicts H$\alpha$ and the left Ly$\alpha$ wing.
  However, the shapes of all lines are reproduced by both models.
  We are investigating these lines since
  the UV lines in this selection represent the two strongest
  ones as measured by \citet{Peacock2019ApJ...886...77P}
  and the H$\alpha$ and \ion{He}{1}~IRT lines
  exhibit discrepancies between the two models.
  The \ion{Na}{1}~D$_2$ line at $5891.6\,\AA$ and the
  bluest \ion{Ca}{2}~IRT line at $8500.4\,\AA$, which have also been used
  in the modeling conducted by \citet[][]{Hintz2020A&A...638A.115H},
  are almost identical in the models M1 and M2 (see Figure~\ref{fig_bis_temp_2b}
  in Appendix~\ref{SEC_App_extra}).

  Finding synthetic spectra that best match the observed lines involves modifying
  the parameters characterizing the densities and temperature rises of the chromosphere and
  transition region. However, significant departures from LTE often
  complicate the behavior of spectral lines with respect to changes in the
  temperatures structure. As discussed by
  \citet{Peacock2019ApJ...871..235P, Peacock2019ApJ...886...77P}, the
  overall UV spectrum is very sensitive to the column mass density at the base of
  the transition region. However, individual lines respond differently, and in
  some cases with opposite trends with respect to changes in the temperature profile.
  For example, the H$\alpha$ line depends strongly on the structure of the upper chromosphere and lower
  transition region while the \ion{Ca}{2}~IRT lines are less affected by these
  atmospheric regions, and vice versa for the structure of the lower
  chromosphere \citep{Hintz2019A&A...623A.136H}.
  However, H$\alpha$ is not independent of the structure of the
  lower chromosphere either, illustrating the difficulty of finding an
  appropriate model that simultaneously reproduces the different line
  observations.

\begin{figure*}[t!]
\includegraphics[width=0.99\textwidth]{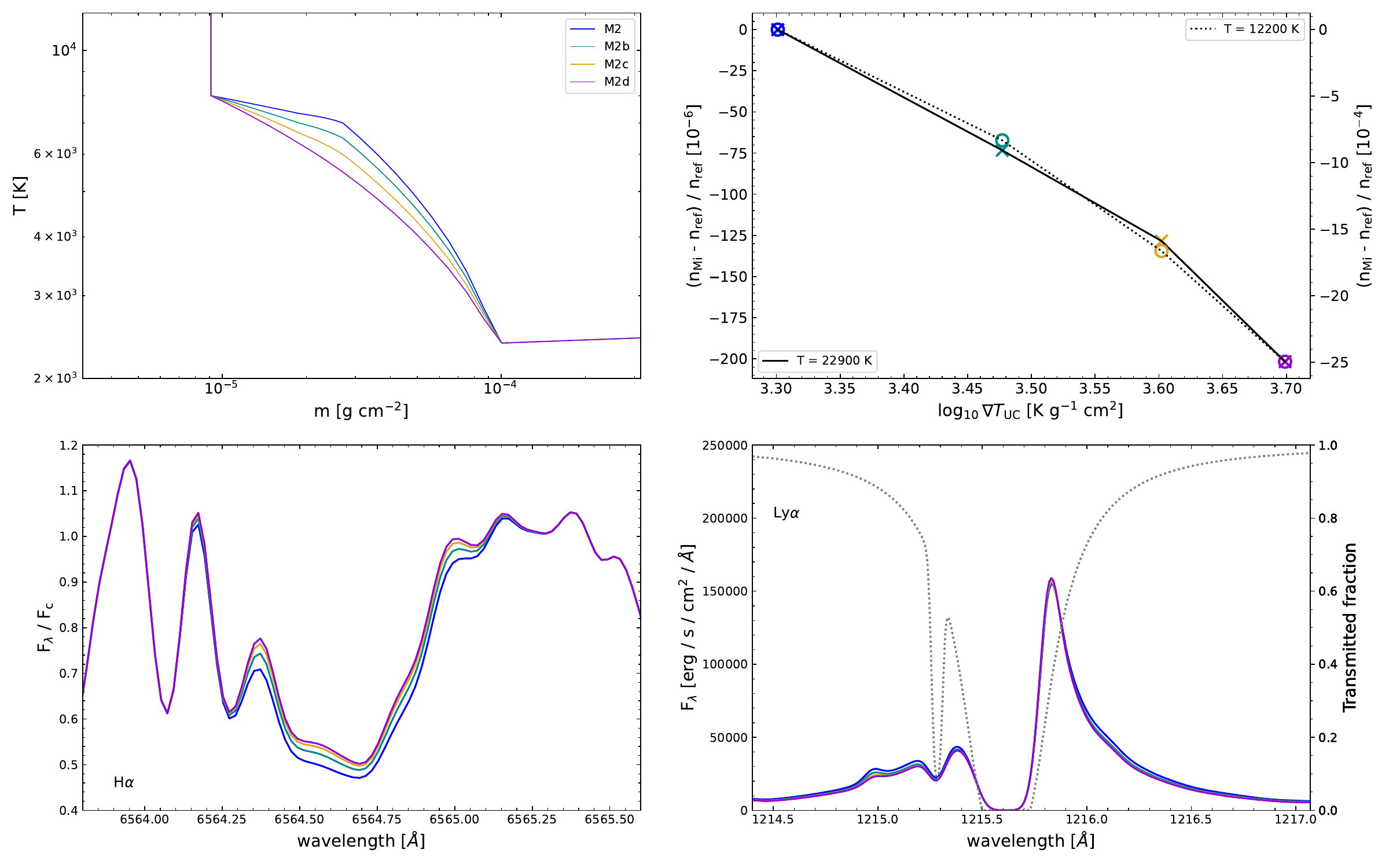}
\caption{
    \textit{Top left:}
    temperature structures of a series of models
    with varying temperature gradients in the
    upper chromosphere ($\nabla T_\mathrm{UC}$).
    \textit{Top right:}
    changes in
    number densities
    of atomic hydrogen
    of models M2b to M2d ($n_\mathrm{Mi}$) relative
    to the reference model M2
    (blue, $n_\mathrm{ref}$)
    at temperatures of 22\,900 (solid line, left axis)
    and 12\,200\,K (dotted line, right axis) from where
    significant contributions to the Ly$\alpha$ and H$\alpha$
    line flux arise, respectively
    (see Figure~\ref{fig_T_struct}, left panel).
    They are plotted against the corresponding
    $\nabla T_\mathrm{UC}$ gradients.
    Colors coincide with the models depicted
    on the left-hand side.
    Circles depict variations in
    number density
    at 12\,200\,K
    and crosses those at 22\,900\,K.
    Increasing $\nabla T_\mathrm{UC}$ from M2 to M2d leads
    to a decrease in both $n_\mathrm{Mi}$ ($T = 12\,200$\,K) and
    $n_\mathrm{Mi}$ ($T = 22\,900$\,K), but
    the effect is significantly stronger
    for $n_\mathrm{Mi}$ ($T = 12\,200$\,K).
    Therefore, H$\alpha$ is more affected by this temperature
    profile parameter than Ly$\alpha$
    (see also Figure~\ref{fig_bis_temp_2c}).
    \textit{Bottom:}
    corresponding model spectra in the
    lines of Ly$\alpha$ (\textit{right},
    including the
    ISM transmission curve)
    and H$\alpha$ (\textit{left}).
    This figure is continued in Figure~\ref{fig_bis_temp_2b}
    in Appendix~\ref{SEC_App_extra} for the
    \ion{C}{4} and \ion{He}{1}~IRT lines.
    }
\label{fig_bis_temp_2}
\end{figure*}

  An important reason for the differences between
  models M1 and M2 shown in Figure~\ref{fig_lines_comp}
  is the position of the transition region in terms of column mass densities
  as introduced in Section~\ref{SEC_Phx_model}.
  The transition region is crucial for all of the lines considered
  in Figure~\ref{fig_lines_comp}.
  Increasing the present activity state, which corresponds to a
  transition region located at higher densities, i.~e., model M2 compared to M1,
  strengthens the line absorption of the H$\alpha$ and \ion{He}{1}~IRT
  \citep[][]{Cram_Mullan_1979ApJ...234..579C,Avrett1994IAUS..154...35A,Hintz2020A&A...638A.115H}.
  Yet the effects of changing the transition region densities
  are more obvious in the UV lines since they are formed at
  higher temperatures than H$\alpha$ and the \ion{He}{1}~IRT.

  Figure~\ref{fig_bis_temp_2} shows a series of models
  with similar temperature profiles but
  varying temperature gradients in the upper chromosphere (UC).
  Starting with model M2,
  we increased the temperature gradient of the upper chromosphere,
  as given by
  \begin{equation} \label{eq_tgraduc}
    \nabla T_\mathrm{UC} = \frac{\mathrm{d} T}{\mathrm{d} \log m} ,
  \end{equation}
  where $T$ is the temperature and $m$ the column mass density.
  In this series, the upper chromosphere starts at $m = 10^{-4.5}$\,g\,cm$^{-2}$
  and the corresponding temperatures vary in steps of
  500\,K from 5500 to 7000\,K (from M2d to M2).
  The corresponding model spectra in the
  H$\alpha$ region illustrate
  a shift of the activity level (from M2d to M2;
  see also the activity measurements displayed in
  Figure~\ref{fig_bis_temp_2c}), i.~e., it
  comes along with an enhancement in the H$\alpha$ line absorption
  (the H$\alpha$ equivalent width increases by 12.5\,\% from M2d to M2).
  The Ly$\alpha$ flux
  also increases, but the line is less affected
  than H$\alpha$
  (the integrated Ly$\alpha$ flux changes by 4.6\,\% from M2d to M2).
  It has a similar effect on both hydrogen lines but is weaker
  than the shift of the transition region to higher densities
  (model M1 to M2, the integrated Ly$\alpha$ flux increases by
  37.1\,\% while the H$\alpha$ equivalent width increases by 15.7\,\%).
  The resulting changes in the \ion{C}{4} and \ion{He}{1}~IRT lines
  when varying $\nabla T_\mathrm{UC}$
  are illustrated in Figure~\ref{fig_bis_temp_2b}.
  Shifting the transition region to higher densities increases
  the strength of both hydrogen lines
  and also of the \ion{C}{4} line,
  while the \ion{He}{1}~IRT absorption deepens
  as seen in Figure~\ref{fig_lines_comp}.
  The \ion{C}{4} line behaves similarly but changes less
  when changing the temperature gradient in the upper chromosphere.
  The \ion{He}{1}~IRT absorption increases with a high-density transition region,
  but weakens when approaching a plateau-like structure in
  the temperature profile of the upper chromosphere.
  On the other hand, the \ion{Na}{1}~D$_2$ and bluest \ion{Ca}{2}~IRT lines
  mostly depend on the lower chromospheric temperature regions
  and are thus just marginally affected when adapting the upper chromospheric parameters.

  The behavior of a line is dependent upon the
  number densities
  at the line formation depths, and thus these
  are useful for illustrating
  what happens when a prescribed atmospheric structure is modified.
  In Figure~\ref{fig_bis_temp_2}, we show
  the number densities of atomic hydrogen at
  two different locations in the transition region indicated by two different
  atmospheric temperatures.
  These are layers from which significant contributions to Ly$\alpha$ and H$\alpha$ originate.
  For this we used the temperatures of 22\,900 and 12\,200\,K
  for Ly$\alpha$ and H$\alpha$, respectively.
  These temperatures correspond to the central temperatures
  from the formation regions indicated in Figure~\ref{fig_T_struct}.
  An increase in the number density
  of atomic hydrogen where
  H$\alpha$ forms is connected with an increase in the H$\alpha$ absorption.
  The change in the
  number density of atomic hydrogen at 12\,200\,K
  from M2 to M2d is one order of
  magnitude larger than is the case
  in the temperature layer from which a significant contribution to Ly$\alpha$ arises.
  The number densities of atomic hydrogen
  at the Ly$\alpha$ and H$\alpha$ formation depths increase
  when shifting the transition region to higher densities as
  well as when establishing a plateau-like profile of the upper chromosphere
  ($\nabla T_\mathrm{UC}$ decreases from M2d to M2).
  However, a shift of the transition region
  to higher densities leads to an increase in the Ly$\alpha$ flux,
  but it does not necessarily lead to the formation
  of the H$\alpha$ line in the same way.
  These findings indicate that the line formation results from
  an interplay between several temperature profile
  properties, such as
  the temperature gradient of the upper chromosphere as well as the onset
  and temperature gradient of the transition region, rather
  than just shifting a prescribed
  temperature structure in terms of column mass densities.

  These illustrations reveal that it is indeed quite difficult to reproduce
  the UV and VIS/NIR lines simultaneously with just one model.
  The UV lines arise from hotter layers in the transition region
  \citep[][]{Linsky2017ARA&A..55..159L} and are strongly
  dependent on its shape and also on the location of the
  temperature minimum \citep[][]{Peacock2019ApJ...871..235P}, while the
  chromospheric densities are important for the lines in the VIS/NIR
  \citep[][]{Hintz2019A&A...623A.136H},
  yet the location of the temperature minimum plays an important role
  in their formation as well.
  At this point it is necessary to account for
  requirements of the different line formation heights
  as well as for the discrepancies arising from
  the complexity of simultaneously modeling UV and
  VIS/NIR lines observed at different times.
  Observations from different time periods
  can be affected by the activity cycle and may require
  a different approach than modeling the considered lines
  with just one single model.

\newpage
\subsection{Linear combination of two models} \label{SEC_lico}
  By analyzing the line strengths of chromospheric indicators
  such as H$\alpha$ in M-dwarf stars it is possible to estimate surface filling factors of
  active regions
  \citep{Giampapa1980SAOSR.389..119G,Giampapa1985ApJ...299..781G}.
  It is common to divide the surface into two components:
  \begin{equation} \label{EQ1}
      F = (1-A) \, F_q + A \, F_a
  \end{equation}
  where $F$, $F_q$, and $F_a$ represent the total,
  quiet, and active line fluxes, respectively, and
  $A$ denotes the fraction of all active regions on the stellar surface
  while the fraction of quiet regions
  is given by the factor $(1-A)$.
  This filling-factor ansatz
  can be approximated by linear combinations of model components
  \citep{Giampapa1985ApJ...299..781G,Ayres2006ApJS..165..618A}.
  This method has already been used in the recent past
  using PHOENIX models by \citet{Hintz2019A&A...623A.136H}
  following the ansatz of Eq.~(\ref{EQ1})
  in order to model active and variable M stars, including GJ~436.
  \citet{Hintz2019A&A...623A.136H} found that
  the filling factor of the active model component increases
  with increasing activity states, but the
  difference in terms of the activity state between inactive and active model
  components decreases for less active stars.

\begin{table}[t]
\caption{
Estimating weighting factors ($w_\mathrm{i}$) in Eqs.~(\ref{EQ2}) and (\ref{EQ3}) from
line errors ($\sigma_\mathrm{i}$) and widths ($c_\mathrm{i}$).
}
\label{tab_gj436_line_measurements}
\begin{center}
\footnotesize
\begin{tabular}{l c c c c}
\hline\hline
 Line & $\lambda_\mathrm{i}$ [$\AA$] & $c_\mathrm{i}$ [$\AA$] & $\sigma_\mathrm{i}$ [$\AA$]$^a$ & $w_\mathrm{i}$ [$\AA^{-2}$] \\
\hline
 Ly${\alpha}$$^b$ & 1215.7   & 3.5  & 2.01 E+4 & 2.47 E-10 \\ 
 \ion{C}{4} &  1548.2  &   0.8 &  2.10 E+3 & 2.26 E-8 \\
 H${\alpha}$ & 6564.6   &   0.9 &  1.46 E-1 & 4.69 E+1 \\
 \ion{He}{1} IRT & 10\,833.3   &  0.9 &  3.36 E-2 & 8.86 E+2 \\
\hline
\end{tabular}
\end{center}
\tablecomments{
$^a$ Line errors correspond to the standard deviations
(within $\lambda_\mathrm{i} \pm c_\mathrm{i}/2$)
depicted in Figure~\ref{fig_lines_comp}.
$^b$ Due to interstellar absorption within the Ly$\alpha$ core
we neglect the inner core region of $1215.7 \pm 0.3\,\AA$
when determining $\sigma_\mathrm{i}$ and
applying the described $\chi^2$-minimization to it.
}
\end{table}

  We linearly combine a model
  capable of reproducing the UV lines and another one that
  reproduces the VIS/NIR lines, aiming at an
  overall match of the investigated lines
  -- models M1 and M2 from
  Figure~\ref{fig_lines_comp}, respectively.
  We use the Ly$\alpha$, \ion{C}{4}, H$\alpha$, and \ion{He}{1}~IRT
  lines for this purpose and we adapt the modified $\chi^2$-minimization
  used by \citet{Hintz2019A&A...623A.136H} in order to look
  for the best combination of the two models.
  Due to the different line widths and
  strengths, we weight the respective line contributions
  when applying the modified $\chi^2$-minimization
  in the wavelength ranges of the lines in order
  to approach an equal treatment of the lines as follows:
  \begin{equation} \label{EQ2}
      \chi^2_\mathrm{m} = \sum_\mathrm{i} w_\mathrm{i} \, \chi^2_\mathrm{i},
  \end{equation}
  where $\chi^2_\mathrm{i}$ corresponds to the sum of the
  quadratic model deviations from the observed flux values within a line $i$.
  The weighting factors $w_\mathrm{i}$ for the lines are determined
  from the standard deviations $\sigma_\mathrm{i}$
  within the widths ($c_\mathrm{i}$) of the respective
  emission and absorption lines:
  \begin{equation} \label{EQ3}
      w_\mathrm{i} = 1 / \sigma_\mathrm{i}^2 .
  \end{equation}
  Table~\ref{tab_gj436_line_measurements} lists the weighting factors,
  $c_\mathrm{i}$, and $\sigma_\mathrm{i}$ for the considered lines.
  With these modified $\chi^2$-values we are able to compare the linear
  combinations among the models and identify our best match.
  Applying these to the $\chi^2_\mathrm{m}$-minimization introduced above
  yields contributions of $(1-A) = 80\,\%$ and $A = 20\,\%$ from
  model M1 and model M2, respectively, to the linear combination.
  Including \ion{Na}{1}~D$_2$ and \ion{Ca}{2}~IRT lines in this
  procedure would only marginally affect the results
  since M1 and M2 look quite similar in these lines.
  The resulting best linear combination would be nearly identical
  to the one without \ion{Na}{1}~D$_2$ and \ion{Ca}{2}~IRT lines;
  the change in the fraction $A$ would be of the order of $\sim1\%$.
  The red model spectrum in Figure~\ref{fig_lines_comp} shows the
  corresponding modeled lines from the linear combination of these two models.
  It exhibits $\chi^2_\mathrm{m} = 45.1$
  compared to $\chi^2_\mathrm{m} = 50.5$ for model M1 only, and
  $\chi^2_\mathrm{m} = 134.4$ for M2, corresponding to an improvement of 10.7\% from M1
  and 66.4\% from M2 for the overall match.
  With this approach we can get an overall match to the observations.
  This means that the H$\alpha$ and \ion{He}{1}~IRT
  cores are matched by the linear combination model
  while we obtain a reproduction of the Ly$\alpha$ and \ion{C}{4} lines as well.

\begin{figure}[t!]
\includegraphics[width=0.49\textwidth]{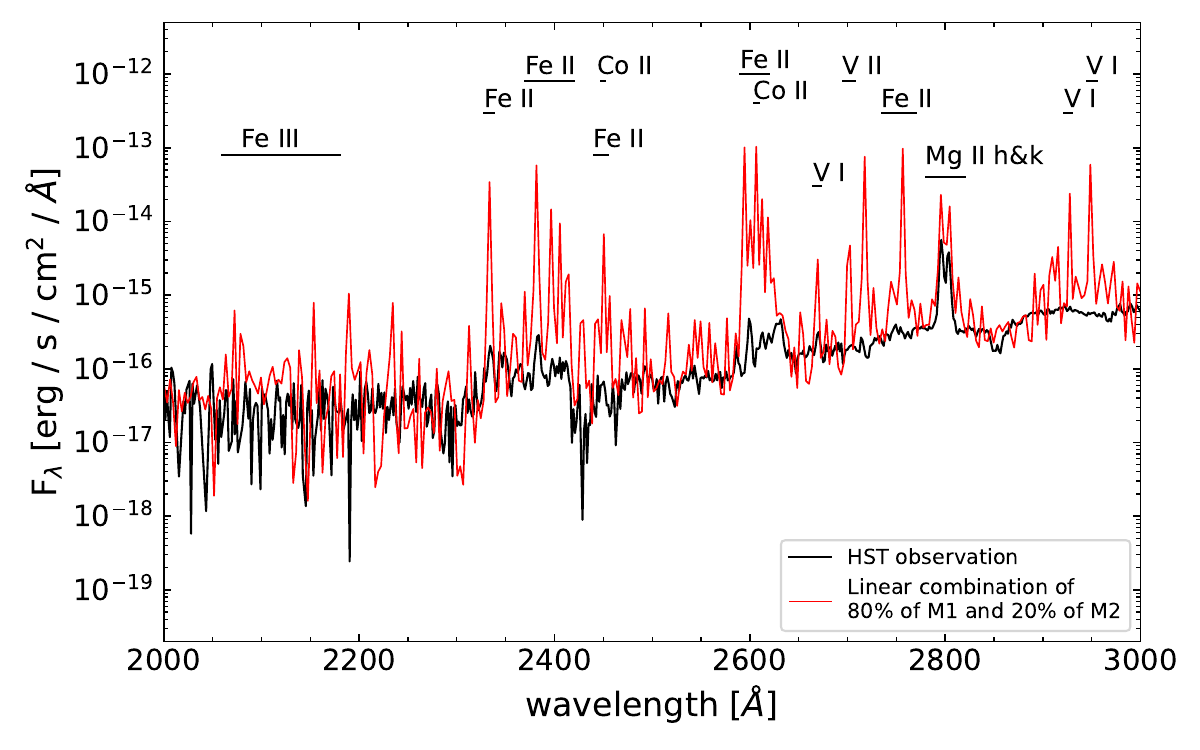}
\caption{
    The linear combination obtained from Section~\ref{SEC_lico}
    for the NUV region compared to an HST STIS observation (G230L grating) of GJ~436.
    The model spectrum is reduced to the resolving power of 1000
    corresponding to the maximum resolution of the STIS G230L grating.
    Here we also indicate a couple of strong lines
    and line bands within this spectral region.
    }
\label{fig_nuv}
\end{figure}

  The near-UV (NUV) HST observation of GJ~436 and the linear combination
  from Figure~\ref{fig_lines_comp} are plotted in Figure~\ref{fig_nuv}.
  In general, the linear combination reproduces the observations
  in the range between $\sim$2000 and 2300\,$\AA$ as well as in the \ion{Mg}{2}~h and k lines.
  We find model deviations from the NUV observation
  redwards of 2300\,$\AA$, where there are some overpredicted \ion{Fe}{2} features
  overlapping with \ion{V}{1}, \ion{V}{2} and \ion{Co}{2} lines, for instance.
  Even though we did not aim at modeling the NUV region,
  we find that the synthetic spectrum from this model
  can also reproduce the NUV observation.

\subsection{Discussion}
  In this paper, we have aimed at finding a model spectrum
  able to reproduce the average state of GJ~436.
  While we used the coadded template spectrum
  of GJ~436 observed with CARMENES over a
  time period from 2016 January to 2016 November
  to investigate its VIS/NIR spectral range,
  we used single HST observations from 2015 June
  to cover the UV region.
  The model study of \citet{Hintz2019A&A...623A.136H}
  included an activity-related investigation
  on CARMENES observations of M dwarfs
  and revealed that GJ~436 is exhibiting
  low activity variations compared
  to four far more active stars in their sample
  -- for instance, the observed H$\alpha$ and \ion{Ca}{2}~IRT lines
  of GJ~436 appeared in pure absorption and
  only showed marginal changes in their line strengths.
  The actual contrast in terms of activity between the two model components
  in the linear combination found here is relatively low,
  for example, compared to the combinations obtained
  for the four active stars investigated in \citet{Hintz2019A&A...623A.136H}.
  So the active component represents the active regions
  of an essentially not very active star,
  which is in agreement with the star's
  measured $\log R'_\mathrm{HK} < -5.1$
  \citep[e.~g.,][]{Fuhrmeister2023A&A...670A..71F}.
  However, we cannot exclude stronger activity-related changes,
  although it has not been observed yet
  and GJ~436 only showed low activity
  in other studies \citep[e.~g.,][]{Lothringer2018AJ....155...66L}.

  For the Ly$\alpha$ line it is even harder
  to make a statement on the variability
  on GJ~436 because interstellar absorption
  considerably suppresses the observed line flux
  \citep[e.~g.,][]{Dring1997ApJ...488..760D,Wood2005ApJS..159..118W},
  and there is still a lack of continuous UV observations of M-dwarf stars.
  From the Sun it is known that Ly$\alpha$ is highly variable over a range
  of more than 30\,\% due to the stellar rotation and activity cycle
  \citep[][]{Vidal-Madjar1975SoPh...40...69V}.
  However, GJ~436 is known to be a bright Ly$\alpha$ emitter
  \citep[][]{Ehrenreich2011A&A...529A..80E}.
  Therefore, it was important to include the Ly$\alpha$ line in our modeling
  to take into account atmospheric heights where high-energy radiation is produced.
  To account for these difficulties we excluded the inner Ly$\alpha$ line core
  in our investigation and we also included a second strong line from the UV region.
  We also compared our final linear combination model spectrum to the
  observed NUV range of GJ~436 in order to verify
  that we obtained a model that also reproduces
  the spectral region between the UV and VIS/NIR lines discussed above.

  As well as the astrophysical deficiencies described above
  there are also model-dependent issues as to why
  the observed lines are not entirely matched by the models.
  Model-related shortcomings may be the
  considered NLTE species set and the element abundances.
  One model with the chosen NLTE set incorporating most of the abundant
  elements already consumes several thousands of CPU hours.
  However, for future modeling purposes and with improving
  computing performances, it will become easier
  to take into account even more species in NLTE.
  Furthermore, future modeling should also test different element abundances.
  Trying to vary element abundances across the model atmosphere,
  which may be reasonable with respect to the
  very hot and low-density regime of the transition region,
  represents another possibility to improve the models.
  Besides, we use semiempirical temperature profiles to approach
  the nature of the chromosphere and lower transition region.
  Future modeling should test temperature structures that account for a balance
  between heat conduction, ambipolar diffusion,
  and radiative losses in order to establish a more
  realistic perspective on the low-density, atmospheric regions
  such as what has been done for solar models by
  \citet{Fontenla1990ApJ...355..700F,Fontenla1991ApJ...377..712F,Fontenla1993ApJ...406..319F}
  using the PANDORA code \citep{Avrett1992ASPC...26..489A},
  but with a comparably restricted NLTE set.

\newpage
\section{Summary and conclusion} \label{SEC_sum_concl}
   In this work, we stated the difficulty of modeling planet-hosting star
   GJ~436 for different spectral features from the UV
   to the VIS/NIR wavelength regions.
   Different line formation temperatures that are separated
   by several thousand Kelvin and optically thick lines
   such as the hydrogen Ly$\alpha$ and H$\alpha$
   lines that form over broad atmospheric regions
   reveal a high complexity when searching for a model
   able to match all considered spectral lines.
   The temperature structure itself
   plays a significant role for the final model spectrum
   when adjusting the structure parameters with regard to the
   different line responses.
   It is known that observed variabilities of
   active low-mass stars
   arise from starspots
   \citep{Reinhold2019A&A...621A..21R}, thus
   it is generally reasonable to account for this by combining
   different models in order to approximate the real
   nature of the stellar surface.
   We applied a linear combination approach
   to simulate the heterogeneity of the stellar surface
   of GJ~436 and we were able to obtain an overall
   match of the UV and VIS/NIR lines.
   The more active model component in the combination
   better matches H$\alpha$ while the more inactive component
   better reproduces \ion{C}{4}.
   The higher activity state was necessary in order to reproduce
   the absorption strength of H$\alpha$.
   The preferred active model component
   contributes about 20\,\% to the linear combination,
   but the main difference between the two components
   is the transition region located at different densities.
   Because of the very different formation regions of the lines that are
   taken into account, we also checked the obtained
   combination in the intermediate range of the NUV.
   Therefore, we conclude that it is sufficient
   to model the M~dwarf GJ~436 with
   a linear combination approach consisting
   of two model components.
   A similar approach should be considered to model
   other stars of this spectral type in the same way.
   Even though GJ~436 exhibits only small activity-related variations,
   we showed the efficiency of the linear combination of two model spectra,
   representing different activity states,
   for the reproduction of spectral features observed at different time periods.
   In general, this is a powerful and very flexible tool to match
   spectral features sensitive to stellar activity when regarding variable stars.

\section*{Acknowledgements}
   Some of the data analyzed in this paper were obtained from the
   Mikulski Archive for Space Telescopes (MAST) at the Space Telescope Science Institute.
   These observations can be accessed via
   \dataset[DOI: 10.17909/qw7n-b890]{https://doi.org/10.17909/qw7n-b890}.
   We thank the anonymous reviewer for the attentive reading
   and suggestions for improvements of the paper.
   We particularly thank Allison Youngblood for
   providing us the ISM transmittance curve for the Ly$\alpha$ line of GJ~436
   from \citet{Youngblood2016ApJ...824..101Y}.
   This work was supported by NASA under the XRP program (grant 80NSSC21K057) and HST- GO-15955.01
   from the Space Telescope Science Institute, which is operated by AURA, Inc., under NASA contract NAS 5-26555.
   This work also benefited from an allocation of computer time from the UA Research Computing High
   Performance Computing (HPC) at the University of Arizona. A portion of the calculations presented
   here were performed at the Hochstleistungsrechenzentrum Nord (HLRN).
   We thank these institutions for a generous allocation of computer time.
   The contributed efforts provided by S.P. are supported by NASA under award number 80GSFC21M0002.
   CARMENES is an instrument at the Centro Astron\'omico Hispano en Andaluc\'ia (CAHA) at Calar Alto (Almer\'{\i}a, Spain),
   operated jointly by the Junta de Andaluc\'ia and the Instituto de Astrof\'isica de Andaluc\'ia (CSIC).
   CARMENES was funded by the Max-Planck-Gesellschaft (MPG),
   the Consejo Superior de Investigaciones Cient\'{\i}ficas (CSIC),
   the Ministerio de Econom\'ia y Competitividad (MINECO) and the European Regional Development Fund (ERDF)
   through projects FICTS-2011-02, ICTS-2017-07-CAHA-4, and CAHA16-CE-3978,
   and the members of the CARMENES Consortium
   (Max-Planck-Institut f\"ur Astronomie,
   Instituto de Astrof\'{\i}sica de Andaluc\'{\i}a,
   Landessternwarte K\"onigstuhl,
   Institut de Ci\`encies de l'Espai,
   Institut f\"ur Astrophysik G\"ottingen,
   Universidad Complutense de Madrid,
   Th\"uringer Landessternwarte Tautenburg,
   Instituto de Astrof\'{\i}sica de Canarias,
   Hamburger Sternwarte,
   Centro de Astrobiolog\'{\i}a and
   Centro Astron\'omico Hispano-Alem\'an),
   with additional contributions by the MINECO,
   the Deutsche Forschungsgemeinschaft (DFG) through the Major Research Instrumentation Programme and
   Research Unit FOR2544 ``Blue Planets around Red Stars'',
   the Klaus Tschira Stiftung,
   the states of Baden-W\"urttemberg and Niedersachsen,
   and by the Junta de Andaluc\'{\i}a.
   We acknowledge financial support from the Agencia Estatal de Investigaci\'on (AEI/10.13039/501100011033)
   of the Ministerio de Ciencia e Innovaci\'on and the ERDF ``A way of making Europe'' through projects
   PID2021-125627OB-C31 and        
   PID2019-109522GB-C5[1:4],    
   and the Centre of Excellence ``Severo Ochoa'' and ``Mar\'ia de Maeztu'' awards to the Instituto de Astrof\'isica de Canarias
   (CEX2019-000920-S), Instituto de Astrof\'isica de Andaluc\'ia (SEV-2017-0709) and Institut de Ci\`encies de l'Espai (CEX2020-001058-M).
   This work was also funded by the Generalitat de Catalunya/CERCA programme and the DFG priority program SPP 1992
   ``Exploring the Diversity of Extrasolar Planets (JE 701/5-1)''.

\bibliographystyle{aastex631}
\bibliography{references}

\begin{appendix}
\counterwithin{figure}{section}

\section{Applying the Ly$\alpha$ ISM transmission curve} \label{SEC_App_ISM}
\begin{figure*}[h]
\begin{center}
\includegraphics[width=0.64\textwidth]{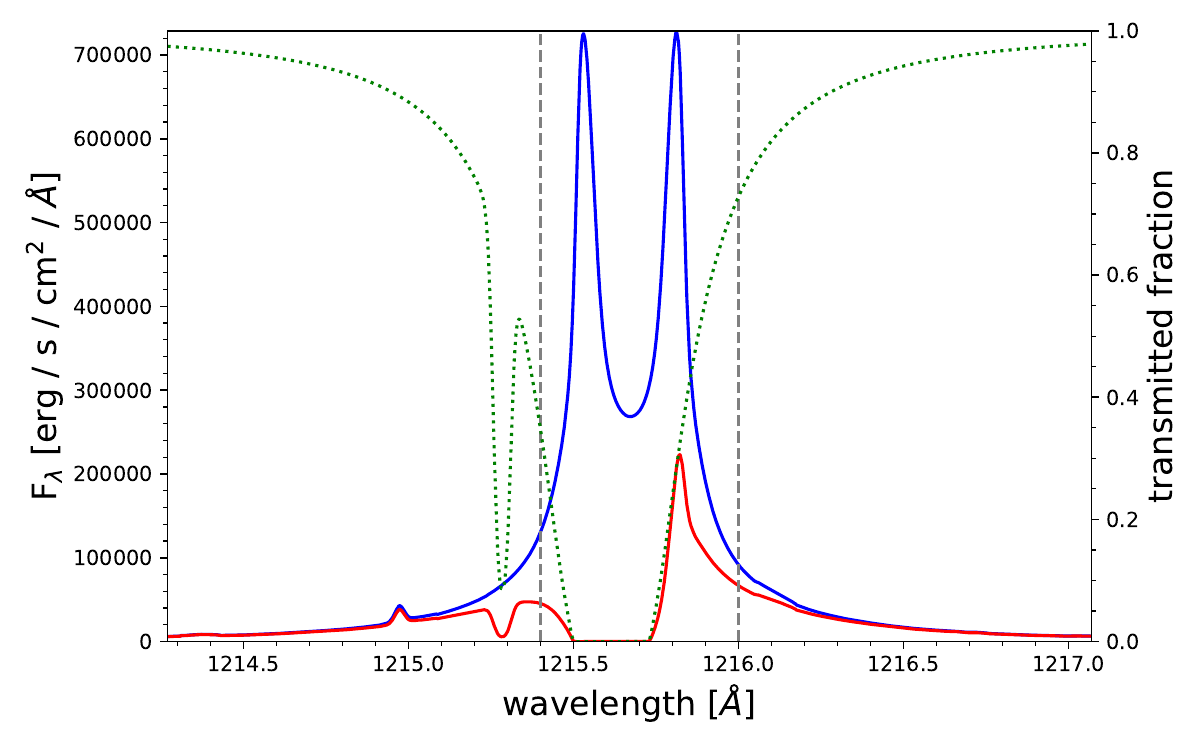}
\caption{
    Application of the Ly$\alpha$ ISM transmission curve (green dotted line)
    for GJ~436 as found by \citet{Youngblood2016ApJ...824..101Y}
    to our model M2 (blue solid line).
    The resulting modeled Ly$\alpha$ line is shown in red.
    We also indicate the Ly$\alpha$ line core area
    we neglected within our modified $\chi^2$-minimization
    by the two dashed vertical lines.
    }
\label{fig_ISM_abs}
\end{center}
\end{figure*}

Supporting Figure~\ref{fig_lines_comp},
we depict in Figure~\ref{fig_ISM_abs} how we apply the ISM transmission curve for GJ~436
using the results obtained from \citet{Youngblood2016ApJ...824..101Y} on model M2.
Onto the original model M2 Ly$\alpha$ spectrum, we deploy the
determined ISM transmission fractions in order to
take into account the interstellar absorption in the
line of sight before accounting for the instrumental broadening
of the HST spectrograph.
Thus, we can directly compare our modeled Ly$\alpha$ line to the observed one.

\section{Modeled line behavior
 when varying chromospheric temperature gradients} \label{SEC_App_extra}
\begin{figure*}[h]
\begin{center}
\includegraphics[width=0.94\textwidth]{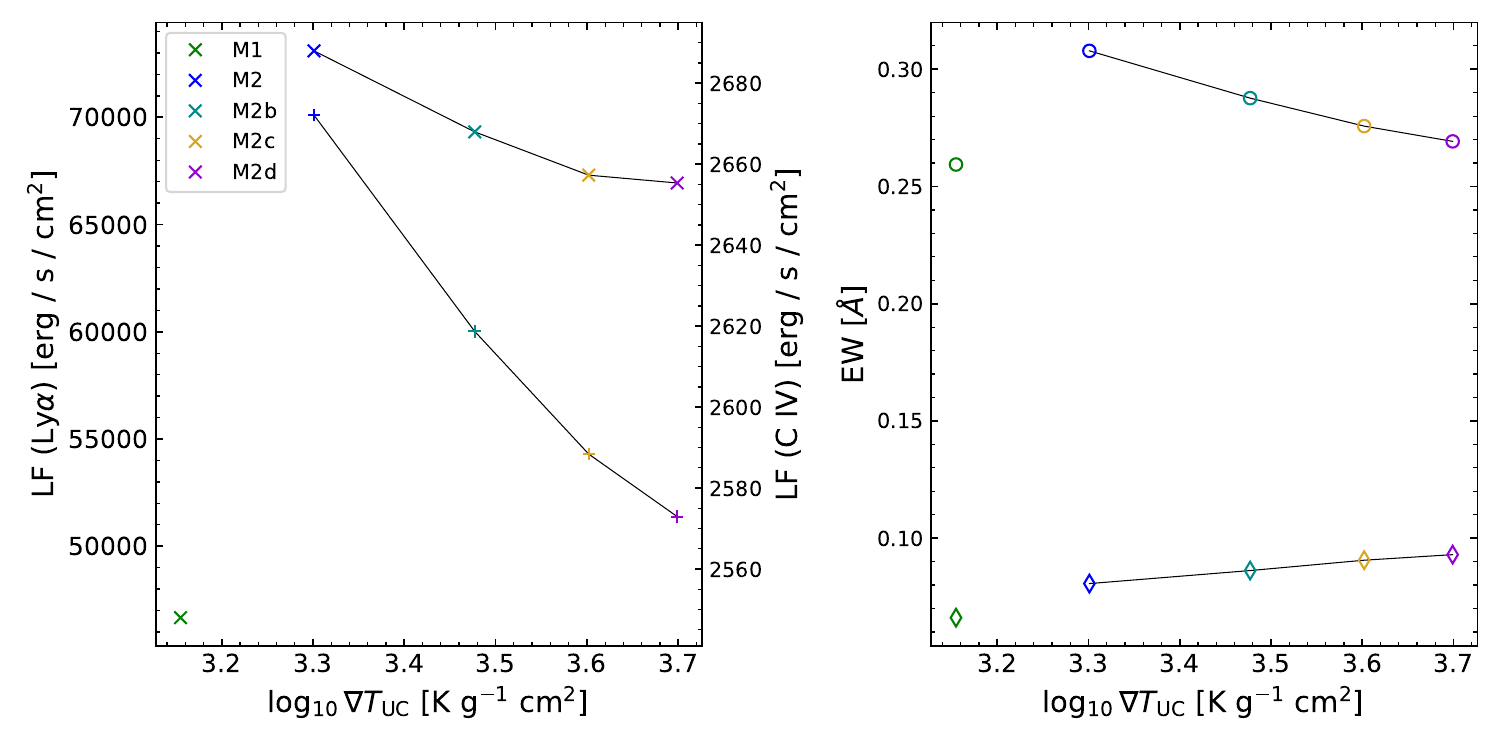}
\caption{
    Activity measurements for model series M2 - M2d as given by
    the integrated line fluxes (LF, \textit{left panel}) of Ly$\alpha$ (crosses) and \ion{C}{4} (pluses),
    as well as the equivalent widths (EW, \textit{right panel})
    of H$\alpha$ (circles) and the \ion{He}{1}~IRT (diamonds)
    for models M2 - M2d as a function of $\nabla T_\mathrm{UC}$.
    The colors of the models coincide with those in Figures~\ref{fig_bis_temp_2} and \ref{fig_bis_temp_2b}.
    We also show the corresponding values of M1 (green markers) except for the
    LF (\ion{C}{4}), which equals to 1251.1\,erg s$^{-1}$ cm$^{-2}$.
    }
\label{fig_bis_temp_2c}
\end{center}
\end{figure*}

\begin{figure*}[h]
\begin{center}
\includegraphics[width=0.93\textwidth]{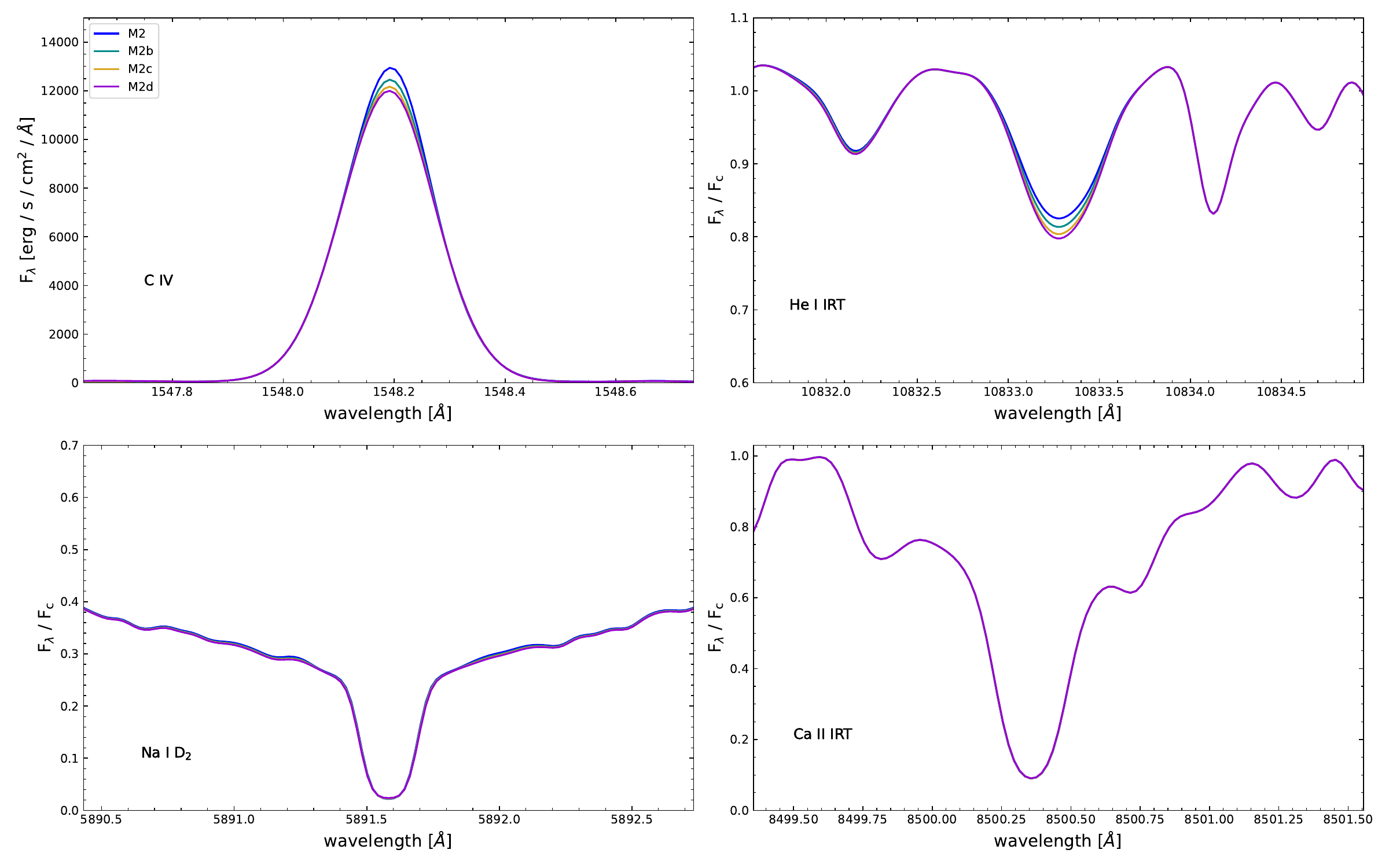}
\caption{
    Figure~\ref{fig_bis_temp_2} continued for
    the \ion{C}{4} (\textit{upper left}),
    \ion{He}{1}~IRT (\textit{upper right}),
    \ion{Na}{1}~D$_2$ (\textit{lower right}),
    and the bluest \ion{Ca}{2}~IRT (\textit{lower right}) lines.
    The colors coincide with those used in Figure~\ref{fig_bis_temp_2}.
    }
\label{fig_bis_temp_2b}
\end{center}
\end{figure*}

 Figure~\ref{fig_bis_temp_2c} shows the corresponding activity measurements
 for the Ly$\alpha$, \ion{C}{4}, H$\alpha$, and \ion{He}{1}~IRT lines
 as given by integrated line fluxes or equivalent widths
 for models M1 and M2 - M2d.
 Furthermore, as a supplement to Figure~\ref{fig_bis_temp_2},
 we continue the displayed line behavior of the models
 in Figure~\ref{fig_bis_temp_2b}
 for the lines of \ion{C}{4}
 and \ion{He}{1}~IRT
 as well as the \ion{Na}{1}~D$_2$, and bluest \ion{Ca}{2}~IRT lines
 when varying the temperature gradient ($\nabla T_\mathrm{UC}$) of the upper chromosphere.

\end{appendix}

\end{document}